# Context-Oriented Web Video Tag Recommendation*

Zhineng Chen[1,2], Juan Cao[1], Yicheng Song[1,2], Junbo Guo[1], Yongdong Zhang[1], Jintao Li[1]
[1]Institute of Computing Technology, Chinese Academy of Sciences, Beijing 100190, China
[2]Graduate School of the Chinese Academy of Sciences, Beijing 100039, China
{chenzhineng, caojuan, songyicheng, guojunbo, zhyd, jtli}@ict.ac.cn

## ABSTRACT
Tag recommendation is a common way to enrich the textual annotation of multimedia contents. However, state-of-the-art recommendation methods are built upon the pair-wised tag relevance, which hardly capture the context of the web video, i.e., *when who* are doing *what* at *where*. In this paper we propose the *context-oriented tag recommendation* (CtextR) approach, which expands tags for web videos under the context-consistent constraint. Given a web video, CtextR first collects the multi-form WWW resources describing the same event with the video, which produce an informative and consistent context; and then, the tag recommendation is conducted based on the obtained context. Experiments on an 80,031 web video collection show CtextR recommends various relevant tags to web videos. Moreover, the enriched tags improve the performance of web video categorization.

## Categories and Subject Descriptors
H.3.1 [**Rich Media**]: Tagging of rich-media data

## General Terms
Algorithms, Performance, Experimentation

## Keywords
Social tagging, context-oriented, web video, tag recommendation

## 1. INTRODUCTION
With the proliferation of web videos available on the WWW, web video tag recommendation – the act of automatically adding relevant tags to web videos – is becoming imperative. As the number of user provided raw tags is limited and they are usually insufficient to characterize the informative web video.

Generally, state-of-the-art methods [1][2] recommend tags using a two-step scheme. First, certain kinds of tag relevance (e.g., tag co-occurrence, tag visual relation, etc.) are mined between each raw tag and potential relevant tag pair; and then, the obtained tag relevance pairs are fused to determine a list of recommending tags. Although encouraging results are reported on web images (e.g. those in [1][2]), trivially applying the scheme to web videos is cumbersome. For example, Fig. 1 gives a video about Pope Bebedict XVI was knocked down by a woman at Christmas mass. By using a method analogous to that in [1], the top 5 recommended tags are "church", "press", "god", "Catholic", and "merry", respectively. The five tags are all meaningless to the video.

We argue, the above mis-recommendation, is attributed to the applied method has not taken into account the context characteris-

*This work was supported by the National Basic Research Program of China (973 Program, 2007CB311100), National Nature Science Foundation of China (60873165, 60802028, 60902090), Beijing New Star Project on Science & Technology (2007B071), Co-building Program of Beijing Municipal Education Commission.

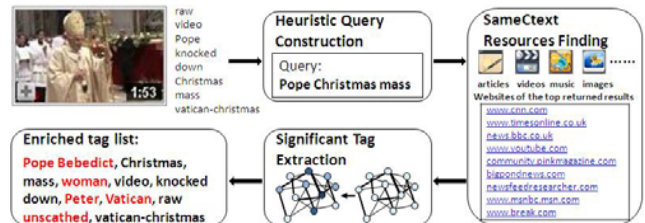
**Figure 1. The illustrative flowchart of CtextR.**

tic of web videos. A web video is generally used to broadcast a dynamic web event, which occurs in a specific context characterized by elements such as who, what, where, when, etc. An isolated raw tag hardly captures such a context, so the tags expanded based on the pair-wised tag relevance are not necessarily in accordance with the video event.

So for web videos, recommending tags in the video-specific event context is essential. But the context-consistent resource in datasets of the same form (e.g., YouTube repository) is limited, which is often insufficient to provide valuable clues for tag expansion. However, the context-consistent multi-form resources on the web supply us with an opportunity to address this deficiency.

In the huge WWW, a same web event is usually covered in many websites in multiple media forms such as videos, news reports, blog articles, etc. For example, the top returned results from Google search include news reports, videos, and user comments from BBS, when using "Pope Christmas mass" as the query, as shown in Fig. 1. These resources provide an informative and consistent context for the web event so we call them the *same-context* (SameCtext) resources in this paper. Therefore, for a web video, if we can collect its SameCtext resources from the web, then relevant tags can be expanded accordingly (see Fig. 1 and Fig. 2).

Motivated by the above observations, in this paper we propose the *context-oriented tag recommendation* (CtextR) approach to recommend tags for web videos. Given a web video, CtextR focuses on finding its SameCtext resources available on the WWW, and expanding and recommending tags from the textual descriptions contributed by these resources. Experimental results on MCG-WEBV [3] show the CtextR approach expands diverse and relevant tags for web videos, which significantly enrich the completeness of the raw tags. In addition, the enriched tags improve the performance of web video categorization.

## 2. THE CTEXTR APPROACH
Given a web video, the CtextR approach works as follows. First, a heuristic is employed to construct a query that captures the context of the video. Then, the query is submitted to Google search engine and the SameCtext resources are identified and collected. The textual descriptions of these resources supply an informative and consistent context for the video. Then, a PageRank-like graph model is applied on the context to discover and rank the keywords from the resources, where the top but untagged ones are added to the video. The whole flowchart is given in Fig. 1.

Table 1. Media form of the collected SameCtext resources

| Form | Number (%) | Form | Number (%) |
|---|---|---|---|
| Videos | 131520 (28.5%) | Music & Radio | 11646 (2.5%) |
| News & Articles | 70189 (15.2%) | Images | 11880 (2.6%) |
| Blogs & BBS | 47239 (10.2%) | Unclassified | 169201 (41.6%) |

## 2.1 Query Construction

Query construction aims at extracting a few keywords from the given video. These keywords characterize the video from several key aspects so they represent the context of the video.

We design a heuristic to implement the extraction. First, the *part of speech* (POS) and *named entities* (NE) attributes of the tags and title words are identified, respectively. Then, the Wikipedia longest match criterion is used to determine tag entities. That is, for example, a video tagged with "Pope, Benedict", "Pope" and "Pope Benedict" are both detected as Wikipedia entities, but only "Pope Benedict" is reserved. We assign a tag entity with a certain POS (NE) if at least one of its tags belongs to this POS (NE).

In the query construction, we give priority to tag entities marked as NE or nouns. To make the query relevant and distinct, each query consists of at least 3 tag entities. For videos containing less than 3 tag entities, the most frequent tag entities in its related videos are added. By this way, query "Pope, Christmas, mass" are constructed for the video given in Fig. 1.

## 2.2 SameCtext Resources Finding

We submit the constructed query to Google and collect the top 10 returned results. Among the results, one is regarded as a SameCtext resource if all query tag entities are presented in its Google generated title or abstract.

## 2.3 Significant Tag Extraction

The textual descriptions attached to the SameCtext resources provide an informative and consistent context for the video. Among them, we believe the significant tags are the ones co-occurred with the query tag entities in multiple resources, so we employ a graph model to discover them. The textual description of the $i$-th SameCtext resource is denoted as $d_i$, which is obtained by combining its Google generated title and abstract. Since not all words are good candidates as the tag, we first apply a syntactic filtering process to $d_i$, where the POS filtering and the Wikipedia longest match are both employed, only the noun, adjective and verb tag entities are reserved. Then we construct an graph $G = (V, E)$, where $V$ is the set of vertices and each vertex is a reserved tag entity, $E$ is the set of edges and its element $e_{i,j}$ is the relation between tag entities $i$ and $j$ designed as

$$e_{i,j} = \sum_{d_k \in D} T_{d_k}(i,j) \quad (1)$$

where $D$ is the textual description collection. If tag entities $i$ and $j$ both appear in $d_k$, then $T_{d_k}(i,j) = 1$, and $T_{d_k}(i,j) = 0$ otherwise.

Based on the obtained graph $G$, the significant score of each tag entity can be deduced. Denoted by $s_k(i)$ the significant score of tag entity $i$ at iteration $k$, the graph model can be formulated as

$$\mathbf{s}_k = \alpha \tilde{\mathbf{E}} \cdot \mathbf{s}_{k-1} + \frac{(1-\alpha)}{|\mathbf{s}|} \mathbf{e} \quad (2)$$

where $\mathbf{s}_k = [s_k(i)]_{|\mathbf{s}| \times 1}$, $\tilde{\mathbf{E}}$ is the row-normalized edge matrix, $\mathbf{e}$ is a vector with all elements equal to 1. $\alpha$ is the damping factor set to 0.85, as in the Google PageRank algorithm.

Eq. 2 has exactly the same form as the PageRank algorithm. So the $\mathbf{s}_k$ in Eq. 2 will converge to $\mathbf{s}_\pi$, the stationary tag entity significant vector, which gives large scores to the tag entities closely linked with many others.

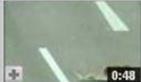

Figure 2: Typical queries and recommended tags.

## 2.4 Tag Recommendation

We rank the tag entities in descending order according to their significant scores, which generates an enriched tag list for the web video. In the list, the ones ranked on the top but not in raw tags are treated as the recommended tags and added to the video.

## 3. EXPERIMENTS AND RESULTS

We conduct the CtextR approach on MCG-WEBV [3], a dataset containing 80,031 representative YouTube videos. We construct queries for 79,043 videos, from which 789,598 resources are collected and 461,768 of them are judged as relevant. We heuristically classify the resources into six forms, according to whether certain pieces appear in their hyperlinks. The result is given in Tab. 1.

As shown in Tab. 1, the queries collect SameCtext resources of multiple media forms, which provide many valuable clues to expand tags for the web videos. Some recommendation examples are given in Fig. 2. From the examples, we can see that the proposed approach recommends various tags to web videos at a relatively low noisy rate.

We perform a tag-based web video categorization experiment on MCG-WEBV to compare the raw tags with the enriched tags. We use the vector space model to build the feature for each video. The raw video category label is used as the ground truth, which assigns each video with one of the 15 YouTube-provided categories. Two SVM classifiers are separately trained and predicted on the two tag sets. The MAP of the enriched tags (0.569) is higher than that of the raw tags (0.521), indicating that tag-related applications have benefit from the proposed CtextR approach.

## 4. CONCLUSION

This paper proposes the CtextR approach to do web video tag recommendation. By leveraging multi-form WWW resources, CtextR successfully recommends relevant tags to web videos. Experimental results on web video categorization demonstrate the feasibility of the context-oriented recommendation idea. In the future, we will incorporate visual features to design a more robust query construction method, which can steadily extract key tag entities no matter how noisy or limited the raw tags are.